\documentclass[letterpaper, 10 pt, conference]{ieeeconf}  

\IEEEoverridecommandlockouts                              
\usepackage[utf8]{inputenc}
\usepackage[T1]{fontenc}
\usepackage{amsmath}
\usepackage{cuted}
\usepackage{lipsum}
\usepackage{optidef}
\usepackage{graphicx}
\usepackage{xcolor}
\usepackage{float}
\usepackage[
backend=biber,
style=numeric-comp,
sorting=none
]{biblatex}
\usepackage{comment}

\title{\LARGE \bf
Optimization of Phase Change Material Integration for Active Cooling Control*
}

\author{Asmaou S. Ouedraogo$^{1}$ and Donald J. Docimo$^{1}$
\thanks{*This work was supported by the National Science Foundation under Award No. 2324707.}
\thanks{$^{1}$Asmaou S. Ouedraogo (aouedrao@ttu.edu) and Donald J. Docimo (donald.docimo@ttu.edu) are with the Department of Mechanical and Aerospace Engineering, Texas Tech University, Lubbock, TX 79404 USA.
}%
}

\begin{document}

\maketitle
\thispagestyle{empty}
\pagestyle{empty}

\begin{abstract}

This paper presents a unified optimization framework for phase change material (PCM) based cooling systems. Thermal management is critical in applications such as photovoltaic (PV) modules, battery packs, and power electronics, where excessive heat reduces performance and lifespan. Designing such systems is challenging because energy dynamics, capacity, heat rejection, and structural constraints must all be considered. Although prior studies have investigated PCM applications and heat transfer enhancement, there are limited efforts that unify such diverse performance objectives through formalized design methods. This paper develops a framework that formulates the PCM design problem using critical energy-based terms, with static and dynamic objectives capturing the PCM physical design and control aspects. Two case studies are used to validate the approach: the first explores passive cooling, and the second implements an active cooling configuration. The results compare the design and control of these systems, showing improvement in individual performance metrics between the two options.

\end{abstract}

\section{INTRODUCTION}
Thermal management has become increasingly critical in all phases of the energy delivery process, from energy generation to energy conversion and storage. Lithium-ion batteries have a narrow safe temperature range for promoting longevity~\cite{chen2023all}, with operation outside of this accelerating aging and risk of thermal runaway \cite{shi2025review}. Photovoltaic (PV) systems, which operate from $20^ \circ \mathrm{C}$ to $75^ \circ \mathrm{C}$ \cite{stritih2016increasing}, show efficiency drops by  0.03-0.06\% per $1^ \circ \mathrm{C}$ rise in temperature~\cite{rahman2015effects}. Elevated temperatures accelerate power electronics degradation by increasing thermal stress, leading to failures like solder fatigue and bond wire lift-off \cite{lai2016study}. As power densities increase and device miniaturization continues, managing heat buildup presents both an opportunity and a challenge. 

PCMs have emerged as a promising thermal management component due to their high latent heat storage capacity, often ranging from 150-250 kJ/kg, allowing them to absorb significant amount of thermal energy \cite{huang2021advances}. Integration of PCM for thermal management has been shown to reduce maximum temperature by  $35^ \circ \mathrm{C}$ in some cases \cite{stritih2016increasing}. However, effective inclusion of PCM in real-world systems is not straightforward, as PCM implementation is subject to inherent limitations that must be carefully considered in design and control. First, the PCM has a finite capacity: once fully melted, it can no longer absorb latent heat and behaves only as a sensible mass, causing device temperature to rise unless excess heat is rejected through a sink. Second, the PCM is recharge dependent: effective cyclic operation requires appropriate conditions between thermal peaks to re-solidify the material. Finally, increasing PCM mass enhances storage capacity but simultaneously adds weight and volume, necessitating a balance achieved through appropriate sizing and active control to maximize utilization. 

Numerous studies have investigated the use of PCM for thermal regulation to address these challenges. A majority of works place the emphasis on passive cooling provided by the PCM to a device or system, evaluating designs based on comparisons and empirical tuning. This includes testing of proposed composite PCM materials~\cite{srivastava2025thermal,ye2022temperature,prabhu2024solar}, altering the physical shape or configuration of the PCM~\cite{nivzetic2021novel,mosaffa2012analytical,zhou2021explicit}, and even including pressure on the PCM to enhance heat transfer~\cite{kim2025cooling}. Some works provide more formal optimization procedures to select PCM material type, the configuration, and size of the PCM, such as for battery cooling applications~\cite{li2025optimisation}. Many of these studies use dynamic conditions to evaluate PCM performance, checking thermal conditions of the PCM or device given a prescribed heat load. Static metrics, such as system mass or volume, are not emphasized despite recognition of the importance for lightweight PVs~\cite{prabhu2024solar} and drones~\cite{pamula2025thermal}. Rare studies do explore the intersection of some aspects of static and dynamic metrics for evaluation, such as minimizing total cooling system mass alongside power consumption for PCMs in active cooling systems~\cite{gulewicz2025set}. There is a need to further promote capabilities in this sector, formalizing design methods that unify, rather than isolate, the key dynamic and static metrics common to PCM cooling systems.

To address the above gap, this work presents an optimization-based design framework to integrate PCMs into actively controlled cooling systems. Using a candidate thermal management system as a starting point, a set of energy-based terms critical to most PCM cooling systems are defined and used to structure the optimization problem. The objective functions to minimize are categorized based on dynamic and static features, easily aggregated using the underlying energy basis. The framework is applied to the candidate system for different potential configurations, facilitating interpretability and comparison of the objective function and design variable values.

The following outlines the structure of the paper: Section II introduces the candidate thermal management system, with the system description and corresponding dynamics. Section III develops the proposed design framework for PCM-based cooling by defining the design variables with related critical energy terms, while distinguishing between the categories of objective functions.  Sections IV and V present case study evaluations, with the former describing the setup and the latter discussing the results. Section VI provides a brief summary of the insights gained in this study.

\section{Candidate Thermal Management System}

This section provides an overview of the candidate active cooling system with thermal energy storage in the form of PCM. For cooling a generic device, component, or load, the system capabilities and potential operating modes are described. Energy balances are presented to represent the thermal dynamics of the system, with approximations made to reduce the model order.

\subsection{System Description}
\begin{figure}[ht]
    \centering
    \includegraphics[width=1\linewidth]{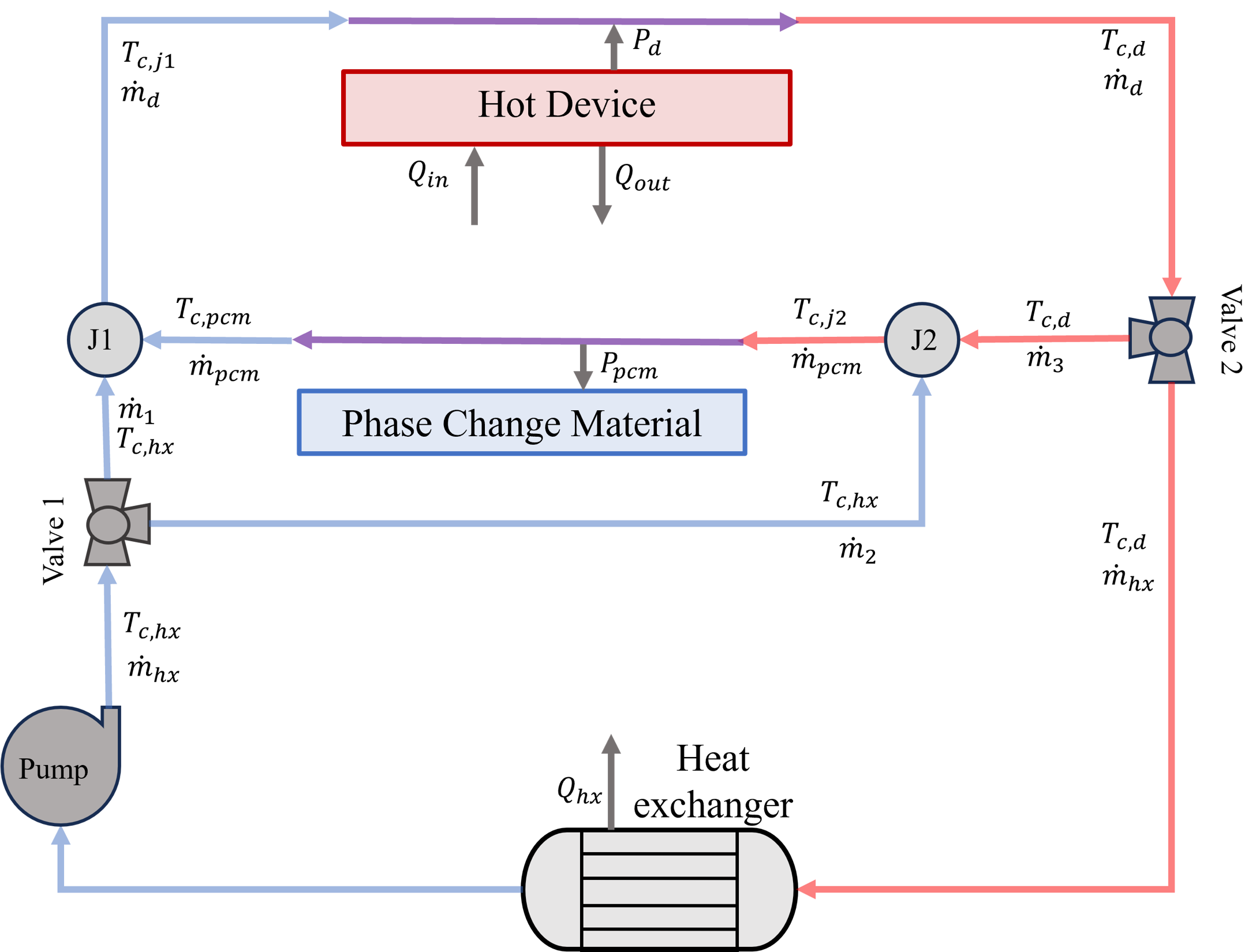}
    \caption{PCM-based active cooling loop configuration, with red for higher coolant temperatures, blue for lower coolant temperatures, and purple for increasing/decreasing coolant temperatures.}
    \label{fig:sys}
\end{figure}

Fig.~\ref{fig:sys} presents a schematic of the proposed hybrid thermal management system with integrated PCM. The hot device, whether a PV, battery, or power converter, requires cooling for short- and long-term operation. The PCM behaves like a thermal battery, absorbing heat at a constant temperature and clipping temperature peaks of the device. The heat exchanger (HX) connects to a secondary active cooling element to provide additional removal of heat as required. These components interact through the coolant circulated using a pump and piping system.

\begin{figure*}[htb!]
    \centering
    \includegraphics[width=1\linewidth]{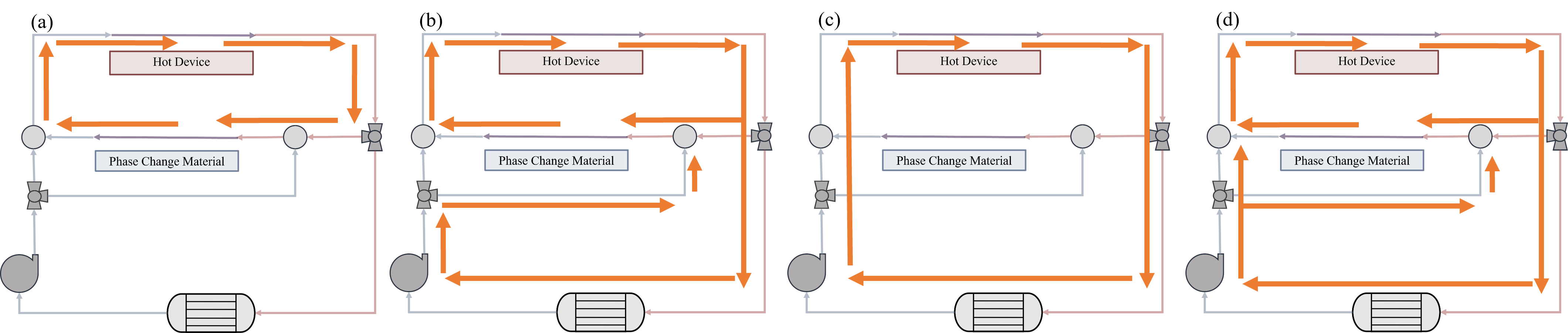}
    \caption{Potential modes for the cooling system.}
    \label{fig:modes}
\end{figure*}

The three-way valves and junctions permit configuration of the thermal management system into alternative operating modes, with examples presented in Fig.~\ref{fig:modes}. The mode of Fig.~\ref{fig:modes}.a circulates the coolant only between the PCM and device, bypassing the heat exchanger. The PCM functions as the sole cooling path, absorbing heat through phase change until its storage capacity is exhausted. The mode of Fig.~\ref{fig:modes}.b routes all coolant through the PCM and device, allowing the PCM to absorb heat from the device while the heat exchanger simultaneously removes heat from the PCM. This enables partial re-solidification and extending PCM availability for subsequent cycles. The mode of Fig.~\ref{fig:modes}.c bypasses the PCM unit, circulating coolant between the hot device and the heat exchanger. The purpose of this mode is to reject heat directly to the environment without engaging a fully discharged (melted) PCM. The mode of Fig.~\ref{fig:modes}.d is the most encompassing, enabling simultaneous latent heat absorption by the PCM and heat rejection through the heat exchanger.

\subsection{System Dynamics}
The dynamic behavior of the thermal management system is mathematically representable using energy balances. The energy balance for the hot device is represented using (\ref{q1}):
\begin{equation}
\label{q1}
C_d\dot{T}_d = Q_{in} - Q_{out} - P_d 
\end{equation}

\noindent where $C_d$ is the device thermal capacitance and $T_d$ is the device temperature. The variable ${Q}_{in}$ captures the heat externally applied to or generated within the device, and ${Q}_{out}$ is similarly defined as the power exiting the device. The term ${Q}_{out}$ excludes cooling heat flow $P_d$ from the coolant, presented in (\ref{q1b}). For this expression, ${T}_{c,d}$ is the temperature of the coolant fluid interacting with the device and $(hA)_{d,c}$ is the product of the effective heat transfer coefficient and area. 
\begin{equation}
\label{q1b}
P_d = (hA)_{d,c}(T_d - T_{c,d}) 
\end{equation}

The energy balance for the PCM is similarly defined in (\ref{q2}), with heat from the PCM to the coolant $P_{pcm}$ defined in (\ref{q2b}). Note that additional expressions can be added to represent the dynamics of a fully melted or fully solid PCM, but are not included here for simplicity:
\begin{equation}
\label{q2}
 C_{pcm}\dot{SOC} =P_{pcm}
\end{equation}
\begin{equation}
\label{q2b}
 P_{pcm} = (hA)_{c,pcm}(T_{c,pcm} - T_m)
\end{equation}

\noindent In these expressions, $C_{pcm}=m_{pcm} L_f$ is the energy storage capacity of the PCM, with mass $m_{pcm}$ and latent heat of fusion $L_f$. The $SOC$ is the melting fraction representing the state of charge of the PCM. The coolant temperature near the PCM is $T_{c,pcm}$, the effective heat transfer coefficient and area is $(hA)_{c,pcm}$, and the melting temperature of the PCM is $T_m$. A lumped capacitance approximation is assumed between the PCM and the coolant, implying negligible internal temperature gradients during the process.

Five additional energy balances are required to model the coolant at different locations. To reduce the model order while preserving the dominant thermal dynamics, the thermal capacitance of the coolant is considered negligible as compared to the device and PCM, setting the rate of change of energy stored in coolant control volumes to zero. This simplification is physically justified when the coolant mass within the control volume is small and its passage time is short due to sufficient flow rates, such that the ratio of coolant thermal capacitance to PCM thermal capacitance remains low. Under this approximation, the coolant behaves as an instantaneous transport medium that transfers energy without contributing to the dynamics. Equations (\ref{q3})-(\ref{q7}) present the energy balances for the coolant in the following locations: at junction 1 (J1), near the hot device, near the heat exchanger, at junction 2 (J2), and near the PCM. These algebraic equations can be used to solve for coolant temperatures $T_{c,j1}$, $T_{c,d}$, $T_{c,hx}$, $T_{c,j2}$, and $T_{c,pcm}$ as outlined in Fig.~\ref{fig:sys}. The parameter $c_p$ is the specific heat capacity of the coolant, and $Q_{hx}$ is the heat rejection from the coolant to the environment through the heat exchanger. 
\begin{equation}
\label{q3}
0 = \dot{m}_{1}c_{p}T_{c,hx} + \dot{m}_{pcm}c_{p}T_{c,pcm} - \dot{m}_{d}c_{p}T_{c,j1}
\end{equation}
\begin{equation}
\label{q4}
0 =P_d + \dot{m}_{d}c_{p}(T_{c,j1}-T_{c,d})
\end{equation}
\begin{equation}
\label{q5}
0 = \dot{m}_{hx}c_{p}(T_{c,d}-T_{c,hx})-Q_{hx}
\end{equation}
\begin{equation}
\label{q6}
0 = \dot{m}_{2}c_{p}T_{c,hx} + \dot{m}_{3}c_{p}T_{c,d} - \dot{m}_{pcm}c_{p}T_{c,j2}
\end{equation}
\begin{equation}
\label{q7}
0 = \dot{m}_{pcm}c_{p}(T_{c,j2} - T_{c,pcm})- P_{pcm}
\end{equation}

The variables $v_1$ and $v_2$ are the throttle commands to split the outgoing fluid for valves 1 and 2, respectively. A value of $v_1=1$ directs all fluid into valve 1 toward junction 2, and a value of $v_2=1$ directs all fluid into valve 2 also toward junction 2. By actively adjusting these flow rates, the system can either enhance or restrict heat transfer depending on the operational requirement. The mass flow rate $\dot{m}_{d}$ is that through the piping near the device, with remaining mass flow rates of Fig.~\ref{fig:sys} determinable using the conservation of mass at steady state, with algebraic relationships defined in (\ref{q8})-(\ref{q12}). 
\begin{equation}
\label{q8}
\dot{m}_{hx} =(1-v_{2})\dot{m}_{d}
\end{equation}
\begin{equation}
\label{q9}
\dot{m}_{pcm} = (v_{1}(1-v_{2})+v_{2})\dot{m}_{d}
\end{equation}
\begin{equation}
\label{q10}
\dot{m}_{1} = (1-v_{1})(1-v_{2})\dot{m}_{d}
\end{equation}
\begin{equation}
\label{q11}
\dot{m}_{2} = v_{1}(1-v_{2})\dot{m}_{d}
\end{equation}
\begin{equation}
\label{q12}
\dot{m}_{3} = v_{2}\dot{m}_{d}
\end{equation}

Equations (\ref{q1})-(\ref{q12}) describe the dynamic model of the cooling system for all modes. With the approximations made, the governing set of energy balance equations is reduced to two differential equations: the device's and the PCM's, covered in (\ref{q1}) and (\ref{q2}), respectively. Together, these components enable a degree of control over the thermal pathway, ensuring that heat is directed efficiently when needed and minimized when it is preferred. 

\section{A Design Framework for PCM-Based Active Cooling}
This section presents the proposed framework to design actively controlled thermal management systems with integrated PCM. As a preliminary step to the framework, a set of energy-based terms are identified as the most critical to describe system performance. This paves the way for definition of the optimization problem's dynamic objective functions, static objective functions, and constraints using these common terms.

\subsection{Design Variables \& Critical Energy-Based Terms}
In the proposed framework, an optimization problem formulation is used to select values of the design variables that minimize an objective function while satisfying equality and inequality constraints. The design variables include those that describe the PCM (e.g., $C_{pcm}$, $T_m$) and the control algorithm parameters. If the desire is to have a controller optimally manage behavior, the latter can be substituting in control input trajectories over time $t$ \big(e.g., $Q_{hx}(t)$, $v_1(t)$, $v_2(t)$\big) as design variables.

What makes the optimization of a cooling system with integrated PCM interesting is the strong connection the prominent objectives, constraints, and design variables all have to energy. Since the system parameters appear through the governing energy balance, a parametric to energy space mapping can be established, allowing the effects of design variables and constraints to be interpreted in terms of their influence on the system energy dynamics. Development of the candidate system model highlighted five critical energy-based terms:
\begin{enumerate}
  \item Energy stored in the hot device, $E_d(t)=C_dT_d(t)$
  \item Energy stored in the PCM, $E_{pcm}(t)=C_{pcm}SOC(t)$
  \item Heat extracted from the hot device, $P_d(t)$
  \item Cooling power provided by the PCM, $P_{pcm}(t)$
  \item Cooling power provided by external sources, such as $Q_{hx}(t)$
\end{enumerate}

\noindent While additional energy-based terms can be defined, these would factor into underlying relationships that feed into the above five terms. These terms are not only critical for evaluation of the candidate system presented in this work alone - evaluation of alternative PCM configurations would still utilize knowledge of $P_d$, for example. The five terms are implemented as the basis of the proposed design framework, with the intention of supporting its applicability to other PCM-based cooling systems.

\subsection{Dynamic \& Static Objective Functions}
The objective functions proposed in the framework seek to capture dynamic- and static-related measures of performance, unifying these through commonality in energy. The total objective function $J_{tot}$ is a combination of the dynamic objective function $J_d$ and the static objective function $J_s$:
\begin{equation}
    \label{q15}
    J_{tot} = w_dJ_d^n + w_sJ_s^n
\end{equation}

\noindent with weights $w_d$ and $w_s$, and $n=1$ traditionally or $n>1$ for compromise programming to mitigate challenges with nonconvex problems \cite{messac2015optimization}.

The dynamic objective function quantifies performance of the evaluated system for a given dynamic profile set of environmental conditions. It is, itself, an aggregation of separate metrics, as presented in (\ref{q16}). This work proposes four internal dynamic objectives, with internal weights $w_{ie}$, $w_{ce}$, $w_{cv,d}$, and $w_{cv,p}$. The $J_{ie}$ function captures input effort, or the integral of cooling power provided by external sources. This relates to the fifth energy-based term of Section III.A. The $J_{ce}$ function describes the integral of heat extracted from the hot device, or the cooling effectiveness, as related to the third energy-based term. As this should be maximized, it is multiplied by $-1$ for minimization problems. The remaining functions, $J_{cv,d}$ and $J_{cv,pcm}$, capture the averaged constraint violations of the stored energy in the device and PCM, respectively. The slack variables, $s_d$ and $s_{pcm}$, are related to the first two energy-based terms, $E_d$ and $E_{pcm}$. A slack variable value at an instance of time is the magnitude of how much energy a component has stored above its upper bound or below its lower bound. In practice, these integrals are calculated using discrete approximations, such as the trapezoidal rule or rectangular method.
\begin{equation}
\label{q16}
\begin{aligned}
&J_{d}=w_{ie}J_{ie}+w_{ce}J_{ce}+ w_{cv,d}J_{cv,d}+w_{cv,p}J_{cv,pcm}\\
&J_{ie}=\int_{0}^{t_f} Q_{hx}(t)\, dt\ \\
&J_{ce}= -\int_{0}^{t_f} P_d(t)\, dt\ \\
&J_{cv,d}= \frac{1}{t_f} \int_{0}^{t_f} s_d(t)\, dt\ \\
&J_{cv,pcm}= \frac{1}{t_f} \int_{0}^{t_f} s_{pcm}(t)\, dt\
\end{aligned}
\end{equation}

The static objective function quantifies performance of the evaluated system for conditions independent of specific dynamic behavior. Similar to $J_d$, $J_s$ is an aggregation of separate metrics with weights $w_m$ and $w_{nom}$:
\begin{equation}
\label{q17}
\begin{aligned}
&J_{s}=w_{m}J_{m}+ w_{nom}J_{nom}\\
&J_{m}=E_{pcm,max} \\
&J_{nom}= -P_{pcm,nom}t_{nom}
\end{aligned}
\end{equation}

\noindent The function $J_m$ is the representative for PCM mass, which should be minimized in most cases. The equivalent to mass in energy-based terms is the maximum value of energy the PCM can store, $E_{pcm,max}=C_{pcm}$, as these are proportional for PCM. The function $J_{nom}$ is the product of the nominal PCM cooling power $P_{pcm,nom}$ and a period of time $t_{nom}$, relating to the fourth energy-based term. This will inherently be dependent on the design variable $T_m$.

As shown by (\ref{q15})-(\ref{q17}), using the basis of the five terms of Section III.A leads to easy unification of dynamic and static objectives. All internal objective functions have units of energy. This facilitates selection of weights for the optimization process, making the weight tuning process more intuitive. Additionally, such representation enhances interpretability and analysis of results, as energy holds practical meaning regarding system behavior and performance. 

\subsection{Constraints}

The constraints of the proposed optimization problem are formulated to ensure that the system remains physically consistent, as well as facilitate calculation of the objective function value. The majority of constraints are defined in terms of energy. Equation~(\ref{q18}) presents the constraints for $k=1,..,N-1$, with $N$ as the horizon for the dynamic profile and $\Delta t$ as the timestep.
\begin{equation}
\label{q18}
    \begin{aligned}
       & [E_{d,1},E_{pcm,1}] = E_{IC}  \\
        & E_{d,k+1} = E_{d,k} + \Delta t (Q_{in,k}-Q_{out,k}-P_{d,k}) \\
        & E_{pcm,k+1} = E_{pcm,k} + \Delta t P_{pcm,k} \\
        & E_{d,k} =  C_d T_{d,k} \\
        & E_{pcm,k} =  C_{pcm} SOC_k \\
        & P_{d,k} = f_d(\cdot) \\
        & P_{pcm,k} = f_{pcm}(\cdot) \\
        & v_{lb} \leq [v_{1,k},v_{2,k}] \leq v_{ub} \\
        & Q_{hx,lb} \leq Q_{hx,k} \leq Q_{hx,ub} \\
        & E_{d,lb} - s_{d,k}\leq E_{d,k} \leq E_{d,ub} + s_{d,k} \\
        & s_{d,k} \geq 0 \\
        & E_{pcm,lb} - s_{pcm,k}\leq E_{pcm,k} \leq E_{pcm,ub} + s_{pcm,k} \\
        & s_{pcm,k} \geq 0 \\
        & C_{pcm,lb} \leq C_{pcm} \leq C_{pcm,ub} \\
        & T_{m,lb} \leq T_{m} \leq T_{m,ub}
    \end{aligned}
\end{equation}

\noindent The first constraint sets the initial condition $E_{IC}$ for the energy storage terms. The discretized, governing energy balances of (\ref{q1}) and (\ref{q2}), with $Q_{in}$ and $Q_{out}$ as disturbances, serve as the second and third constraints. The fourth and fifth constraints relate energy stored to internal quantities for the components. The power flow relationships, shown in the sixth and seventh constraints, contain generic functions $f_d$ and $f_{pcm}$. These can be the same as (\ref{q1b}) and (\ref{q2b}), respectively, with algebraic solutions for $T_{c,d}$ and $T_{c,pcm}$. Alternatively, these can express more advanced internal relationships for a PCM cooling system, including additional dynamics. The eighth line of constraints describes the bounds of the value control inputs, $v_{lb}$ for lower and $v_{ub}$ for upper, and the ninth is the equivalent for the controlled heat exchanger cooling power. Constraints ten through thirteen determine the slack variable values, with $E_{d,lb}$ and $E_{pcm,lb}$ as the lower bounds for the energy stored, and $E_{d,ub}$ and $E_{pcm,ub}$ as the upper bounds for the energy stored. The final two constraints place lower ($C_{pcm,lb}$, $T_{m,lb}$) and upper  ($C_{pcm,ub}$, $T_{m,ub}$) limits on the PCM design variables, with this list expanded as necessary.

\subsection{Solving the Optimization Problem}
The optimization problem is to minimize $J_{tot}$ as described by (\ref{q15})-(\ref{q17}) with constraints (\ref{q18}) for $k=1,..,N-1$ by iteratively selecting values for the decision variables. The decision variables include the PCM design variables, $C_{pcm}$ and $T_m$, and the control input sequences for $Q_{hx,k}$, $v_{1,k}$, and $v_{2,k}$. The decision variables also include $E_{d,k}$, $E_{pcm,k}$, $T_{d,k}$, $SOC_{k}$, $P_{d,k}$, $P_{pcm,k}$, $s_{d,k}$, and $s_{pcm,k}$. Numerical solvers can be used to determine these values. With the constraints being bilinear or nonlinear depending on the cooling system underlying model, careful selection of the appropriate solver is required.

\section{Case Study Setup}
This section sets up the conditions for application of the framework to optimize the candidate system. The hot device for the case studies is a PV module - the cooling system is implemented to increase daily energy production by reducing the operating temperature. The energy equations for PV thermal dynamics are well established~\cite{jones2001thermal,ouedraogo2024model}, being used to define the heat and power flows for (\ref{q1}):
\begin{equation}
\label{q19}
\begin{aligned}
Q_{in}&= \alpha A_s G\\
Q_{out}&= h_{\infty} A_s (T_d - T_{\infty}) + P_E
\end{aligned}
\end{equation}

The power in is a function of absorptivity $\alpha=0.7$, surface area $A_s=0.8~\mathrm{m^2}$, and irradiation $G$. The power out is dependent on two factors: the natural cooling with heat transfer coefficient $h_{\infty}=13.39~\mathrm{W/m^2/^\circ C}$ and ambient temperature $T_{\infty}$, and the electrical power produced $P_E$. The PV is approximated as 20\% efficient from $Q_{in}$ to $P_E$. The thermal capacitance $C_d=4580~\mathrm{J/^\circ C}$ as defined in~\cite{jones2001thermal}.

The functions $f_d$ and $f_{pcm}$ are those defined in (\ref{q1b}) and (\ref{q2b}), with $T_{c,d}$ and $T_{c,pcm}$ determinable by solving the algebraic expressions of (\ref{q3})-(\ref{q12}) analytically. The effective heat transfer coefficients are $(hA)_{d,c}={21.42~\mathrm{W/^\circ C}}$ and $(hA)_{c,pcm}={21.42~\mathrm{W/^\circ C}}$. The coolant parameters are $\dot{m}_d=1.794~\mathrm{kg/s}$ and $c_p=1370~\mathrm{J/kg/^\circ C}$ \cite{karimi2013thermal}.

\begin{figure}[htb!]
    \centering
    \includegraphics[width=1 \linewidth]{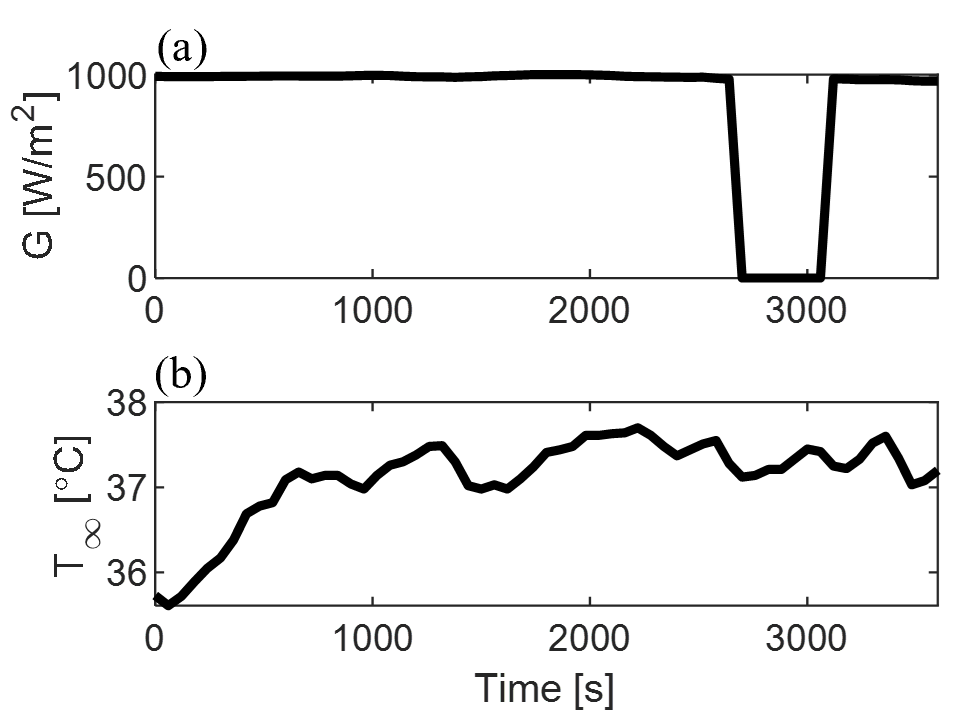}
    \caption{(a) Irradiance and (b) ambient temperature profiles.}
    \label{fig:dist}
\end{figure}

The dynamic profiles for evaluation are presented in Fig.~\ref{fig:dist}, drawn from \cite{Nreldata}. This captures the system operating during the peak hour of the day, where thermal loads on the PV are largest. A drop in irradiation is added near the end of the window to simulate cloud coverage. Initially, the PV device is at a temperature of $T_{d}=35 \mathrm{^\circ C}$ and the PCM is at $SOC=0.5$. The optimizer parameter values include: $\Delta t=60~\mathrm{s}$, $E_d$ bound between $45800~\mathrm{J}$ and $229000~\mathrm{J}$ (corresponding to $10 \mathrm{^\circ C}$ to $50 \mathrm{^\circ C}$ for $T_d$), $E_{pcm}$ between 0 and $C_{pcm}$, $v_1$ and $v_2$ between 0 and 1, $Q_{hx}$ between 0 and 100 W, $C_{pcm}$ between $5 \cdot10^5~\mathrm{J}$ and  $6 \cdot10^6~\mathrm{J}$, and $T_m$ between $20 \mathrm{^\circ C}$ and $50 \mathrm{^\circ C}$. Internal weights found in (\ref{q16})-(\ref{q17}) are set to 1, with the exception of $w_{nom}=0$. The weights $w_d$ and $w_s$ vary for each case study, and $n=1$.  A multi-start initialization strategy was used, where the optimization was initialized from multiple starting points. The simulation and optimization are performed using the MATLAB function \texttt{fmincon} with a 2.5GHz and 32 GB RAM desktop computer. During the simulation using MATLAB's \texttt{fmincon} solver, the optimization settings were specified as a maximum number of function evaluations of $10^8$ (MaxFunctionEvaluations) , a maximum of $3\times10^5$ iterations (MaxIteration), a step tolerance of $10^{-6}$ (StepTolerance), and a constraint tolerance of $10^{-6}$ (ConstraintTolerance). 

\begin{figure*}[htb!]
    \centering
    \includegraphics[width=1 \textwidth]{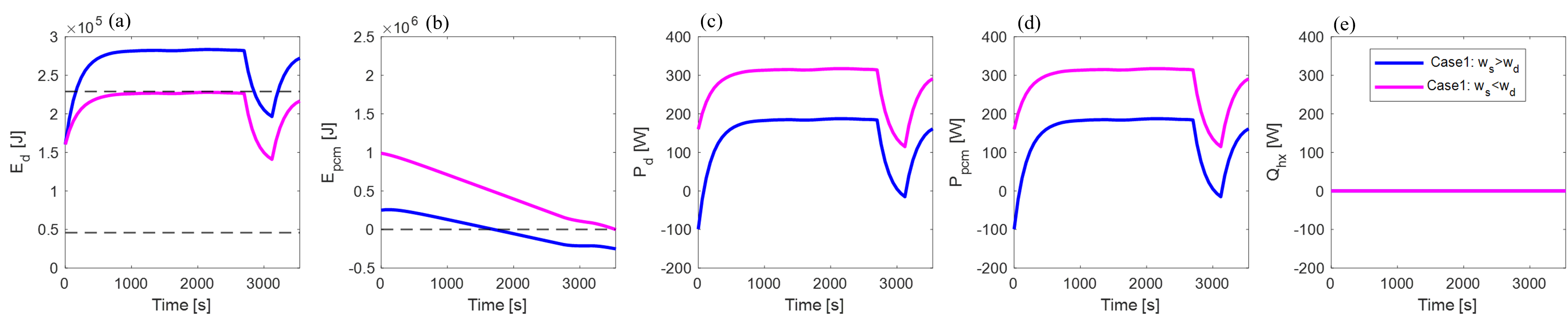}
    \caption{(a) Device energy, (b) PCM energy, (c) heat rejected by the device, (d) heat absorbed by the PCM, and (e) heat rejected to the environment for case study 1 with passive cooling.}
    \label{fig:passive}
\end{figure*}

\section{Results}
This section uses the proposed framework to determine cooling system designs and control through optimization and analyze the decisions made. Two case studies are considered. The first reflects design of a PCM system with passive cooling, and the second provides one option for active cooling system design. Each case is evaluated twice: once weighing static objectives heavily, and once weighing dynamic objectives heavily.

\subsection{Case Study 1: PCM Design for Passive Cooling}
For case study 1, the inputs for the valves are set as constants: $v_1=0$ and $v_2=1$. This puts the cooling system in the configuration of Fig.~\ref{fig:modes}.a, where the heat exchanger is bypassed and the system uses the PCM for passive cooling. Two extremes are considered for analysis. The first emphasizes static objectives by setting $w_s=100$ and $w_d=1$. The second does the opposite, with $w_s=1$ and $w_d=100$.

Fig.~\ref{fig:passive} presents the trajectories over time for $E_d$, $E_{pcm}$, $P_d$, $P_{pcm}$, and $Q_{hx}$. When $w_s>w_d$, $C_{pcm}=C_{pcm,lb}$ is selected. This is expected, as the mass of the PCM is favored to be minimized to shrink $J_{tot}$. There is a tradeoff: with the dynamic objectives deprioritized, the lower bound of $E_{pcm}$ is violated, indicating the PCM melts in the prescribed dynamic conditions. Notably, $E_d$ violates its upper bound as well. This is not solely due to the PCM melting, but rather the optimizer selecting $T_m=44.3^ \circ \mathrm{C}$. While this may seem counterintuitive at first, lower $T_m$ values will increase $P_{pcm}$ and thus melt the PCM faster. This would increase the violation of the $E_{pcm}$ bounds by greater magnitudes, increasing $J_{tot}$.

The alternative scenario, when $w_s<w_d$, yields $C_{pcm}=1.98\cdot 10^6 ~\mathrm{J}$, closer to the upper bound. In this, the PCM is sized large enough to avoid significant violation of the energy bounds, permitting $T_m=T_{m,lb}$. Comparison of the two weighted objectives highlights the tradeoffs of PCM design objective terms. The framework provides insight into how to tune weights: for this case study, designs that balance $J_s$ and $J_d$ would be easier to identify by tuning $w_d$ over $w_s$, due to the former's impact on energy bound violations.

\subsection{Case Study 2: PCM and Control Input Optimization}
\begin{figure*}[htb!]
    \centering
    \includegraphics[width=1 \textwidth]{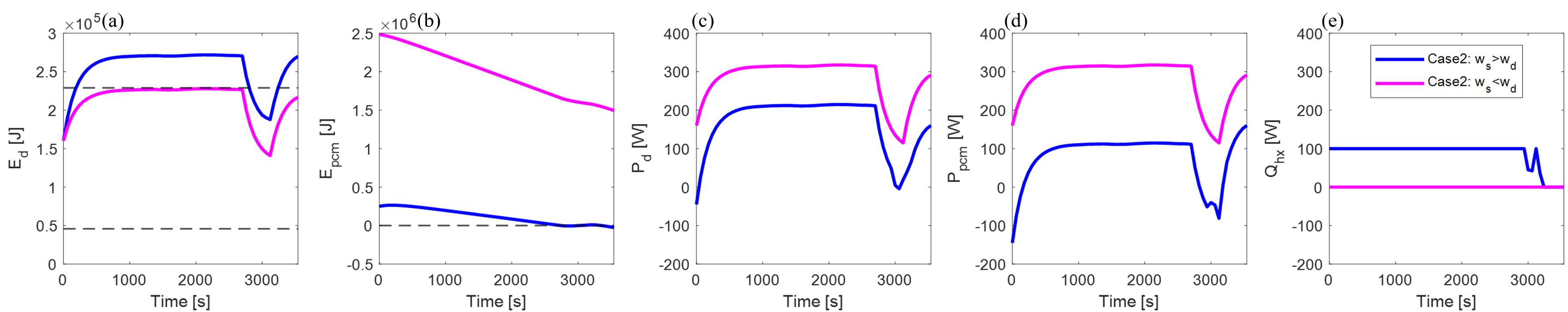}
    \caption{(a) Device energy, (b) PCM energy, (c) heat rejected by the device, (d) heat absorbed by the PCM, and (e) heat rejected to the environment for case study 2 with active cooling.}
    \label{fig:active}
\end{figure*}

For case study 2, the valve inputs are set to $v_1=1$ and $v_2=0.5$, fixing the cooling system in the configuration of Fig.~\ref{fig:modes}.b. All coolant passes through the PCM to the PV, with some rerouted to the heat exchanger for active cooling by controlling $Q_{hx}$. Once again, the scenarios of $w_s>w_d$ and $w_s<w_d$ are explored, using the same values as with case study 1. Fig.~\ref{fig:active} presents the trajectories of the energy-based terms for the given dynamic conditions.

When $w_s>w_d$, $C_{pcm}=C_{pcm,lb}$ and $T_m=43.8^ \circ \mathrm{C}$, consistent with expectations. The primary objective becomes reducing energy storage capacity of the PCM (and thus mass). While penalties for exceeding energy storage bound violations are low, so is the penalty for using $Q_{hx}$. The design process selects a profile of $Q_{hx}$ shown in Fig.~\ref{fig:active}.e: external cooling runs at $Q_{hx,ub}$ until the drop in irradiation, after which it bounces around briefly before going to $Q_{hx,lb}$. This extra cooling is used to reduce melting of the PCM, with $E_{pcm}$ only slightly going under its lower bound.

When $w_s<w_d$, $C_{pcm}=4.97\cdot 10^6 ~\mathrm{J}$ and $T_m=T_{m,lb}$ are selected by the optimization algorithm. With a larger $w_d$, use of $Q_{hx}$ is undesirable, as well as energy storage bound violations. As such, $C_{pcm}$ is increased in size as compared to the $w_s>w_d$ scenario. The trajectory of $E_{pcm}$ in Fig.~\ref{fig:active}.b indicates $C_{pcm}$ may be oversized, as the lower bound is not approached. This suggests that the total objective function is sufficiently numerically flat for the selected weights. Rebalancing of the internal weights for the dynamic objectives can mitigate this, with reductions in $C_{pcm}$ enabled by penalizing use of $Q_{hx}$ less. Such weight tuning is made intuitive by the proposed design framework due to the common units of all objective functions.

\begin{table*}[ht]
    \centering
    \caption{Comparison of case studies 1 and 2 for $w_s=100$ and $w_d=1$}
    \begin{tabular}{ccccccc}
         &  $J_{ie}$&  $J_{ce}$&  $J_{cv,d}$&  $J_{cv,pcm}$&  $J_{m}$& $J_{tot}$\\
         \hline
         Case Study 1 (CS1)&  $9.70\cdot 10^{-16} 
$&  $-5.10\cdot 10^5$&  $3.92\cdot 10^4$&  $8.04\cdot 10^4$&  $5.00\cdot 10^5$& $4.97\cdot 10^7$\\
         Case Study 2 (CS2)&  $3.10\cdot 10^5$&  $-5.94\cdot 10^5$&  $3.03\cdot 10^4$&  $1.32\cdot 10^3$&  $5.00\cdot 10^5$& $4.97\cdot 10^7$\\
  CS1/CS2& $3.21\cdot10^{-21}$& $0.860$& $1.29$& $61.0$& $1.00$&$1.00$\\
    \end{tabular}
    \label{tab:case1}
\end{table*}

\begin{table*}[ht]
    \centering
    \caption{Comparison of case studies 1 and 2 for $w_s=1$ and $w_d=100$}
    \begin{tabular}{ccccccc}
         &  $J_{ie}$&  $J_{ce}$&  $J_{cv,d}$&  $J_{cv,pcm}$&  $J_{m}$& $J_{tot}$\\
         \hline
         Case Study 1 (CS1)&  $1.68\cdot 10^{-4} 
$&  $-9.93\cdot 10^5$&  $1.46\cdot 10^{-4}$&  $1.46\cdot 10^2$&  $1.96\cdot 10^6$& $4.97\cdot 10^7$\\
         Case Study 2 (CS2)&  $4.75\cdot 10^{-5}$&  $-9.93\cdot 10^5$&  $1.51\cdot 10^1$&  $3.33\cdot 10^1$&  $4.97\cdot 10^6$& $4.97\cdot 10^7$\\
 CS1/CS2& $3.53$& $1.00$& $9.64\cdot 10^{-6}$& $4.38$& $0.390$&$1.00$\\
    \end{tabular}
    \label{tab:case2}
\end{table*}

Table~\ref{tab:case1} presents the values of all internal objective functions for $w_s>w_d$ of both cases, and {Table~\ref{tab:case2} presents the same for $w_s<w_d$. Integration of the PCM with active cooling through an external source provides tangible benefits. This is exemplified by the comparison presented in the bottom row of Table~\ref{tab:case1}, with values over 1 indicating benefits provided by active cooling. Due to its negative sign, the exception to this is $J_{ce}$, where values under 1 indicate benefits. While the total objective function is effectively identical between the two cases, there are substantial differences between the internal dynamic objective function values. The most significant difference is between the $J_{cv,pcm}$ values across the different case studies, which is obtained by computing the integral of the energy bound violations over the simulation. The  design for passive cooling violates the PCM  energy bound by a magnitude of 60X more than the active cooling. This is enabled through an increase in input effort $J_{ie}$. While this is a fringe scenario for comparison, weights can be refined to determine alternative solutions. These results illustrate the abilities of PCMs combined with active cooling control.

\section{CONCLUSIONS}
This paper introduces an optimization framework for PCM-based component cooling that integrates energy formulations with static and dynamic objectives. By explicitly defining decision variables, such as melting temperature and control input values, and by enforcing constraints through energy balances and slack variables, the framework enables selection of PCM properties across diverse thermal management applications. The case studies demonstrate that when prioritizing static objectives, such as PCM mass, both passive and active cooling configurations increase melting temperature to reduce energy bound violations for the PCM. However, through control of external resources,  the active cooling configuration's energy bound violations are less than 2\% of those of the passive cooling's. The framework advances cooling design by bridging PCM physical features and control with feasible considerations, providing new insights into the balance between efficiency and thermal control. Its energy-based formulation is broadly applicable to other domains, such as battery systems and power electronics, highlighting its potential for wider adoption.

\addtolength{\textheight}{0cm}






\printbibliography

\end{document}